\documentclass[graybox]{svmult}

\usepackage{mathptmx}       
\usepackage{helvet}         
\usepackage{courier}        
\usepackage{type1cm}        
\usepackage{makeidx}         
\usepackage{graphicx}        
\usepackage{multicol}        
\usepackage[bottom]{footmisc}

\makeindex             

\begin{document}

\title*{Constructing a Cepheid Period $p$-Factor Relation from Static Model Stellar Atmospheres}
\author{Hilding R. Neilson, Chow-Choong Ngeow, and John B. Lester}
\institute{Hilding R. Neilson \at Argelander Institute for Astronomy, Auf dem Huegel 71, 53121 Bonn, Germany, \email{hneilson@astro.uni-bonn.de}
\and Chow-Choong Ngeow \at National Central University, Jhongli City, Taoyuan County 32001, Taiwan (R.O.C.) 
\and John B. Lester \at University of Toronto Mississauga, Mississauga, Ontario, L5L 1C6, Canada}
%
%
\maketitle

\abstract*{One of the largest uncertainties for using the Baade-Wesselink method to measure Cepheid distances is the value of the projection factor (p-factor). However, p-factors measured using the IRSB technique and from hydrodynamic models disagree. In this work, we compute spherically-symmetric static model stellar atmospheres and predict  a period p-factor relation.}

\abstract{One of the largest uncertainties for using the Baade-Wesselink method to measure Cepheid distances is the value of the projection factor (p-factor). However, p-factors measured using the IRSB technique and from hydrodynamic models disagree. In this work, we compute spherically-symmetric static model stellar atmospheres and predict  a period p-factor relation.}

\section{Introduction}
\label{sec:1}
The Cepheid period-luminosity (PL) relation (also called the Leavitt Law \cite{Leavitt1908}) is a powerful tool for cosmological and extragalactic distance measurements. However, the power of the PL relation is limited by its calibration, i.e. independent measurements of Galactic and Large Magellanic Cloud Cepheid distances. One such method for measuring Cepheid distances is the Baade-Wesselink method (\cite{Baade1926, Wesselink1946}).

However, the Baade-Wesselink method depends on the ratio of the observed radial velocity and the pulsation velocity defined as the projection factor or p-factor. The geometric p-factor is defined as
$p \equiv \int I(\mu) \mu d\mu/\int I(\mu) \mu^2 d\mu$,
and  is still the one of the greatest sources of uncertainty in calibrating the PL relation. The variable $\mu$ is the cosine of the angle between a point on the stellar disk and the disk center, while $I(\mu)$ is the intensity at that point.

We compute model atmospheres using the SAtlas code (\cite{Lester2008}), hwere the geometry can be modeled as either spherically-symmetric or plane-parallel.
Models  are computed as a function of pulsation period for a range $\log P = 0.4 - 2$, assuming a bolometric PL relation (\cite{Turner2010}), a period-radius relation (\cite{Gieren1999}) and a period-gravity relation (\cite{Kovacs2000}) to provide input parameters for the code.  From the models, limb-darkening profiles and p-factors are computed for BVRIH and K-bands.

%

\section{Period Projection-Factor}
\label{sec:3}
\begin{figure}[t]
\sidecaption
\includegraphics[scale=0.9]{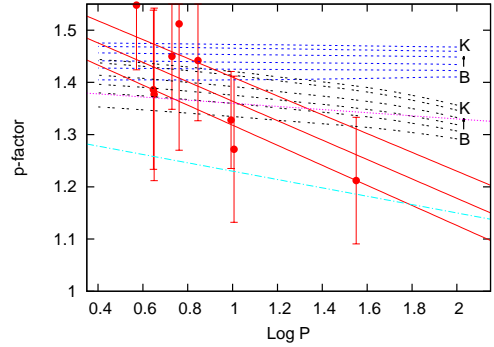}
%
%
\caption{Period-p-factor relations from plane-parallel (blue dashed lines), and spherically- symmetric model atmospheres (black double-dashed line) for the six wavebands. Also plotted are  relations from \cite{Hindsley1986} (magenta dotted line), \cite{Nardetto2009} (pale blue dot-dashed line) and \cite{Storm2011a} with errors (red solid lines). The points are p-factors for a number of Galactic Cepheids (\cite{Storm2011a}).}
\label{fig:pfactor}       
\end{figure}

 The p-factors are plotted in Fig.~\ref{fig:pfactor} along with period-projection factor relations from other sources, as well as p-factors for Galactic Cepheids determined using infrared surface brightness technique and HST astrometry (\cite{Storm2011a}).

Our predicted relation agrees with  results based on the IRSB technique and HST parallaxes (\cite{Storm2011a}) but not with the results from hydrodynamic modeling (\cite{Nardetto2009}).  We also predict a non-linear period p-factor relation, even though our models are derived from linear relations between period and luminosity, radius and gravity.


\begin{acknowledgement}
HRN acknowledges funding from the Alexander von Humboldt Foundation.
\end{acknowledgement}
%

%
%
%
%


\end{document}